\documentclass[twocolumn,aps,prc,superscriptaddress,showpacs]{revtex4}
\usepackage{amsmath,bm}
\usepackage{graphicx}

\begin{document}

\draft
\title{Parallel momentum distribution of the $^{28}$Si fragments from  $^{29}$P}
\thanks{This work was supported  partially
 by the Major State Basic Research Development Program under
 Contract No G200077400,  the Chinese
Academy of Sciences Grant for the Distinguished Young Scholars of
National Natural Science Foundation of China (NNSFC) under Grant
No. 19725521, NNSFC under Grant Nos 10135030 and 10328509 and the
Shanghai Phosphor Program Under Contract Number 03 QA 14066.}
\author{ WEI Yi-Bin}
\affiliation{Shanghai Institute of Applied Physics, Chinese
Academy of Sciences, P.O. Box 800-204, Shanghai 201800}
\affiliation{Graduate School of the Chinese Academy of Sciences}
\author{MA Yu-Gang} \thanks{Email: ygma@sinr.ac.cn}
  \affiliation{Shanghai Institute of Applied Physics, Chinese Academy of Sciences, P.O. Box 800-204,
Shanghai 201800}
\author{CAI Xiang-Zhou}\thanks{Email: caixz@sinr.ac.cn}
\author{ZHONG Chen}
\author{CHEN Jin-Gen}
\author{ZHANG Hu-Yong}
\author{FANG De-Qing}
\author{WANG Kun}
\author{MA Guo-Liang}
\author{GUO Wei}
\author{TIAN Wen-Dong}
\author{SHEN Wen-Qing}
  \affiliation{Shanghai Institute of Applied Physics, Chinese Academy of Sciences, P.O. Box 800-204,
Shanghai 201800}
\author{ZHAN Wen-Long}
\author{XIAO Guo-Qing}
\author{XU Hu-Shan}
\author{SUN Zhi-Yu}
\author{LI Jia-Xing}
\author{GUO Zhong-Yan}
\author{WANG Meng}
\author{CHEN Zhi-Qiang}
\author{HU Zheng-Guo}
\author{CHEN Li-Xin}
\author{LI Chen}
\author{MAO Rui-Shi}
\author{BAI Jie}
\affiliation{Institute of Modern Physics, Chinese Academy of
Sciences, Lanzhou 730000}


\begin{abstract}
Distribution of the parallel momentum of $^{28}$Si fragments from
the breakup of 30.7 MeV/nucleon $^{29}$P has been measured on C
targets. The distribution has the FWHM with the value of 110.5
$\pm$ 23.5 MeV/c which is consistent quantitatively with Galuber
model calculation assuming by a valence proton in $^{29}$P. The
density distribution is also predicted by  Skyrme-Hartree-Fock
calculation. Results show that there might exist the proton-skin
structure in $^{29}$P.
\end{abstract}
　　

 \maketitle

Studies have been performed for many years for exotic nuclei with
the increasing availability of radioactive nuclear beam around the
world. The exotic structures of nuclei refer to the weakly bound
systems that display the very diffuse surface of nearly pure
nucleonic matter at densities far below that of normal nuclear
matter. The nucleus $^{11}$Li is the first observed case as a
neutron halo nucleus \cite{Tanihata1}, and other nuclei that are
considered to have neutron halo are $^{6}He$ \cite{Kobayashi},
$^{14}Be$ \cite{Suzuki,Tanihata_sum} and etc. Such kind of halo
nucleus provide good venue to investigate the weakly bound quantum
system, which was not easily accessible before. Nuclear
interactions in a low-density nuclear matter and loosely bound
three-body interactions are among the current themes of interest
related to these nuclei. In addition to the neutron halo, several
experiments suggest the existence of proton halo nuclei. However,
it is not yet fully established. More attentions are needed on the
proton halo structure. Recently, the nuclei including $^{8}B$
\cite{Tanihata1}, $^{12}N$ \cite{Warner} and $^{23}Al$ \cite{Cai},
which have the very low one-proton separated energy, are
considered to have single proton halo structure. Compared with the
neutron halo, proton halo is difficult to be formed because of the
Coulomb repulsion interaction, but it is still an experimental
observable phenomenon.

The methods which are used to study the halo structure have been
explored and extended in recent years. For the size of the nuclei,
the methods of the total reaction cross section and the one
nucleon knocked-out cross section are widely
used\cite{Shen,Ma0,Ma,Ma1}. From which, the sensitive probes to
the abnormal structure of the nuclei, namely the radius and the
density distribution can be deduced. However,  to some nuclei,
which have the larger cross section compared with the neighbored
isotopes, may be not the candidates of the halo ones. This
attributes to the extended density distribution of the core of the
nucleus, which would induce the larger cross section
\cite{Saint,Bazin}. Therefore, it is necessary to find a suitable
approach to explore the internal structure of the nuclei.  The
momentum distribution of the nucleus fragment could accord with
such requirement. In a highly simplified picture, the longitudinal
momentum distribution of fragment from the breakup of a loosely
bound projectile directly reflects the internal momentum
distribution of the valence nucleon and hence the square of the
Fourier transform of its wave function. In the previous works, it
was shown that the width of the momentum distribution can be
understood in terms of the Fermi motion or a temperature
corresponding to the nuclear binding energy
\cite{Goldhaber,Fujita}, and the momentum distribution of the
valence-nucleon in a beam nucleus can be deduced from the observed
momentum distribution of the fragment by taken the evaporation
correction \cite{ZPA,Ma_width}. Thus, halo formation in such
loosely bound nucleus can be investigated by measuring the
momentum distribution of the fragment from a breakup reaction. The
wide spatial dispersion  of a halo nucleon translates into a
narrow momentum distribution. Many practical works have been done
in the measurements of the momentum distribution widely both
theoretically and experimentally
\cite{Tanihata1,Kobayashi,Guimaraes}. It has been reliable while
searching for the exotic structure of the nuclei. Very recently,
the momentum correlation function among the nucleons of the
neutron-rich nuclei has also been investigated and it seems to be
sensitive to the binding energy per nucleon and single neutron
separation energy \cite{Wei,Wei2}.

In the previous works, the study of the proton-halo structure
mainly focuses on the light proton-rich nuclei, whose mass numbers
are not more than 25. It is interesting to investigate the
characters of the nuclei with the mass number around 30. Navin et
al. performed an experiment to study the structure of the
$^{26-28}$P \cite{Navin}. Some investigations of the density
distribution of these P isotopes have also been
done\cite{Ren1,Ren2}. Judging from the small single proton
separation ($S_p$) energy (0.14 MeV, 0.9 MeV, 2.065 MeV,
respectively) for $^{26,27,28}$P \cite{Audi}, it is easy to detach
the valence-proton from the core. This favors to form the proton
halo structure, which has been approved through their narrow
fragment momentum distributions. Considering the small $S_p$ of
$^{29}$P (2.748 MeV),  it should be interesting to measure the
momentum distribution of the one-proton removal fragment of
$^{29}$P. In this Letter, we report  the experimental measurement
and model analysis for the parallel momentum distribution of
$^{28}Si$ from the break-up of $^{29}$P on $C$ target.


\begin{figure}
\includegraphics[scale=0.6]{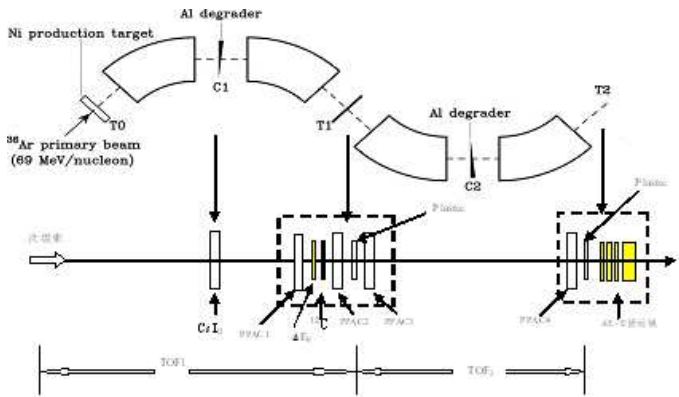}
\caption{\footnotesize The experiment set-up for the measurement
of parallel momentum distribution in RIBLL. For details see
texts.} \label{Fig1}
\end{figure}

The experiment  was performed at the Radioactive Ion Beam Line in
Lanzhou (RIBLL). $^{29}$P secondary beam was produced by the
fragmentation of a 69 MeV/nucleon $^{36}$Ar primary beam on  Be
target which was delivered by the Heavy Ion Research Facility in
Lanzhou (HIRFL). A maximum intensity of 60 enA for $^{36}$Ar was
used to produce the secondary nuclear fragments. As shown in Fig.
1, the production target ($Be$) was mounted on the target box
(T0). Before the target there is one $Si$ detector which could
give the identification of the nuclei. The angular acceptance was
6.5 msr, and the fragments emitted between -2.61$^{o}$ and
2.61$^{o}$ could be transmitted through the spectrometer. The good
collection is obtained in a cone around these angles because of
the strongly focus of the projectile fragments. After T0, RIBLL
can be used as a doubly achromatic magnetic spectrometer where
nuclei produced in the reaction are separated and selected by
means of magnetic rigidity (B$\rho$) and energy degrader
($\Delta$E). The momentum acceptance $\Delta$p/p of the magnetic
spectrometer was set at about 10$\%$

The variable slits and energy degraders, which were made by Al
foils and mounted on curved frames, were placed at the dispersive
focal plane C1 and C2. The time of flight (TOF) of the fragments
was measured by two scintillator detectors installed at the first
(T1) and second (T2) achromatic focal planes with a flight
distance of 16.8m. The measurement of momentum distribution was
done using the TOF. The $Si$ detector before $C$ target is 300
$\mu$m. A telescope was installed at T2, which consisted of three
transmission $Si$ surface barrier detectors followed by a CsI (T1)
crystal and gave the energy losses ($\Delta$E's) and total energy
of the reaction products, where the identification of nuclei could
be completed. The thickness of  three $Si$ detectors were 150, 150
and 1000 $\mu$m, respectively, and the energy resolutions of the
Si detectors were not greater than 0.8$\%$. It is possible to
identify the reaction products and get the momentum distribution
combining TOF and $\Delta$E \cite{Caiexp,Fang_CPL,fangexp}.

With the B$\rho$-$\Delta$E selection of RIBLL, the secondary beam
of $^{29}$P can be identified  before $C$ target. The PPACs will
ensure the position of $^{29}$P to focus on the center of the
target. After the target, the reaction products will be selected
by the B$\rho$-$\Delta$E again. After the flight during the TOF2,
the significative events will be accepted by the final three $Si$
detectors. With the TOF2, the fragments of the reaction will be
achieved. Thus, the calibration of the TOF and $Si$ detectors are
very important.  The good linear relationship between the channel
number and the flight time (ns) have been obtained. From the
flight time of the nuclei, we could get the longitudinal
(parallel) momentum distribution. Also, in this experiment, the
angular distribution can be deduced from the position signals
obtained through three PPACs before and after the C target.

\begin{figure}
\includegraphics[scale=0.75]{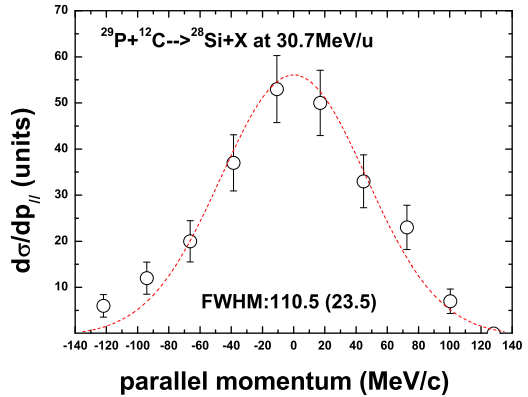}
\caption{\footnotesize The momentum distribution data from
one-proton removal in $^{29}$P on $Be$ target at 30.7 MeV/nucleon.
The dashed line is the Glauber calculation result. The error bars
are only the statistical. } \label{Fi2}
\end{figure}

\begin{figure}
\includegraphics[scale=0.75]{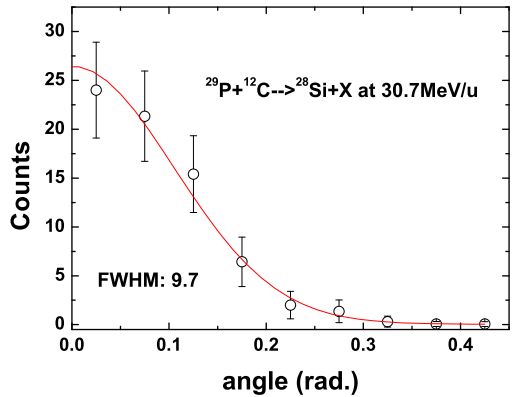}
\caption{\footnotesize The angular distribution data from
one-proton removal in $^{29}$P on $Be$ target at 30.7 MeV/nucleon.
The solid line is the fit with single Gaussian distribution.}
\label{Fig3}
\end{figure}

Fig. 2 shows the experimental result of the parallel momentum
distribution and Fig. 3 displays the angular distribution of the
$^{28}$Si fragment in $^{29}$P. The width (FWHM) of momentum
distribution was found to be 110.5 $\pm$ 23.5 MeV/c and that of
scattering angular distribution is 9.2$^{o}$ which is obtained
from the one-Gaussian fit.

As we know, the lower nucleon separation energy favors to form the
exotic structure. For the light proton halo nuclei which have been
discovered, the one proton separation energy is smaller than that
of the neighboring nuclei among its isotopes. The proton most far
away from the core due to its small separation energy will appear
in the more extended region which is defined halo. For the
candidates of proton halo nuclei, $^{8}$B, $^{14}$N, $^{17}$Ne and
$^{23}$Al all have the smaller one proton separation energies. For
proton-rich nuclei, it is possible to form the proton halo
structure just similar with the neutron halo existing in the
neutron-rich nuclei. For $P$-isotope, the number of proton is 15.
$^{26-29}$P have the more number of protons than that of neutrons.
Considering the small one proton separation energy, one could
imagine the possibility of the proton halo structure existing in
the proton-rich nuclei of $P$-isotope. For $^{26,27}$P nuclei, the
experiment results have achieved to indicate their halo structure
\cite{Navin,Liu,Fang_CPL,fangexp}. But for $^{28}$P, it is not
sure for the existence of the exotic structure \cite{fangexp}.
Compared with the separation energy of $^{28}$P, it is more
difficult to form the halo structure for $^{29}$P. But it is still
interesting to investigate its structure for the normal or
proton-skin character.

  We attempt to obtain the theoretical results using  the Glauber model, which
   can give the reasonable results of the
  fragment momentum distribution. The Glauber model is a  microscopic reaction theory  based on the
  eikonal approximation and on the bare nucleon-nucleon
  interaction. It is now a standard tool to calculate the reaction
  of a weakly bound nucleus. The observed interaction cross sections can be related to
   the wave functions of these nuclei through this model and also one can obtain
   information on the structure of these exotic isotopes. The
   accurate nuclear wave functions play the very important role in
   the reaction simulation. Varga et al developed the Monte
   Carlo integration in Glauber model and some good results have
   been obtained \cite{Varga}. Abu-Ibrahim et al. developed this model and some reasonable
   results have been achieved compared with the experiment \cite{B.Abu}.
  We took his code in our calculation.
  More elaborated treatment beyond the optical-limit approximation
  is worked out for the valence-nucleon part by using the Monte
  Carlo quadrature with the Metropolis algorithm. This is suitable
  for reactions involving those nuclei which have spatially extended and low
  density distribution as in a halo nucleus.

  In the model, the density of the projectile is treated as
  two parts: the core and the valence nucleon density distributions.
  The core density is assumed to be given by a combination of Gaussians:
 \begin{equation}
 \rho(r)=\Sigma c_{i}e^{-a_{i}r^{2}}
 \end{equation}
 Also, the density $\rho$ is normalized to the mass number of a
 nucleus, $\int$d$\textbf{r}$$\rho$(r)=A. The valence-nucleon
 density distribution will be generated through the model with the
typical parameters including the separated energy, the angular
momentum value and so on. The density of the target is treated
similar with that of the core of the projectile. Thus, the
valence-nucleon character will be included in the Glauber
calculation and the results show that such consideration is
reliable. More details about Glauber model could be found in
Ref.~\texttt{\texttt{}} \cite{B.Abu}.

\begin{figure}
\vspace{-0.2truein}
\includegraphics[scale=0.4]{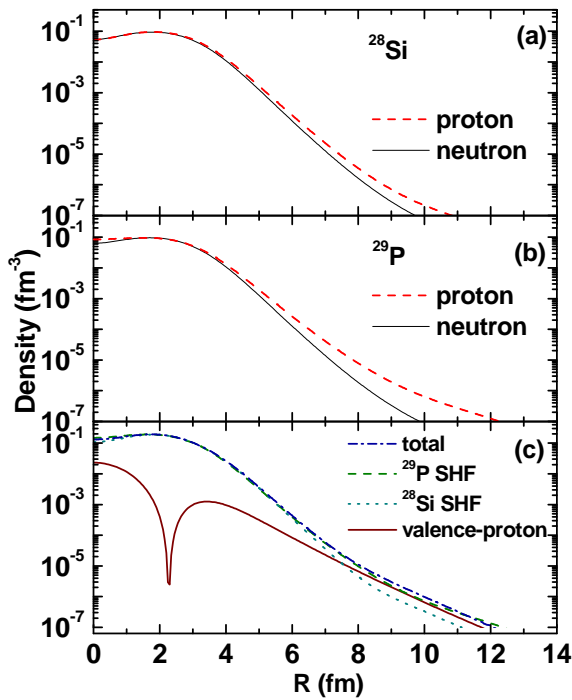}
\vspace{-0.1truein}
 \caption{\footnotesize The calculated density
distribution of protons and neutrons for $^{28}$Si (a) and
$^{29}$P (b) with SHF model. In (c), the valence-proton density
distribution is from Glauber model while the density distribution
of  $^{28}$Si and $^{29}$P are from SHF model; the total one is
the sum of the valence-proton density and the density of
$^{28}$Si.} \label{Fig4}
\end{figure}

Using such a model, we calculated the reaction of the
30.7MeV/nucleon $^{29}$P on  $^{12}$C target. The parallel
momentum distribution of the $^{28}$Si fragment after one-proton
removal from  $^{29}$P has been obtained. The calculated FWHM is
91 MeV/c and the result is the dashed line shown in Fig. 2. In our
calculation, the valence proton density distribution of $^{29}$P
is decided by the separation energy, the angular momentum value
and the node of the density distribution of the valence proton.
The core of the projectile and the target density distributions
are fitted with five gauss function according to the calculated
results with Skyrme-Hartree-Fock(SHF) method. Thus, the parallel
momentum distribution of $^{29}$P fragmentation is self-consistent
in experiment and theory.

In Ref.~\cite{Zahar}, the fragment angular distribution will give
the almost same character with the momentum distribution. The
angular distribution indicates the scattering character of the
projectile. Compared with the neutron-halo nuclei \cite{Marques},
the angular distribution could also describe the structure in the
proton-rich nuclei. Fig. 3 shows our experimental results. The
peak of the angular distribution of $^{28}$Si fragment shows the
scattering angular during the reaction.  From the value of the
angular distribution, it is hard to say the existence of the
exotic halo structure in $^{29}$P. But the proton-skin structure
might exist.

In order to understand the structure of $^{29}$P further, it is
more clear to describe $^{29}$P from the sight of nucleon orbits.
For $^{29}P$, the last proton orbit is 1s1/2, and the other 14
protons and 14 neutrons form the structure that similar with that
of $^{28}$Si. The single-particle energies of protons in level
1d5/2 and 2s1/2 for $^{29}$P are, respectively, 11.585MeV and
2.748MeV. These indicate that $^{29}$P can  be approximately
considered as a $^{28}$Si core plus one proton because the
valence-proton is weakly bound. One could use some models to study
such nucleus through the density distribution. The SHF is one of
the suitable models which can give the reasonable nucleus
structure \cite{Ren1,Ren2}. We calculated the density
distributions of $^{29}$P and $^{28}$Si shown in Fig. 4. The
dashed lines in (a) and (b) are the proton density distributions.
From the figure one find that the density distribution of neutrons
in $^{28}$Si and $^{29}$P are very similar, but the density
distribution of protons in $^{29}$P extends farther than that of
$^{28}$Si. It displays that the radii of proton and matter of
$^{29}$P are a little larger than those of $^{28}$Si. This
displays the weakly unbound status of the last proton of $^{29}$P,
namely the proton-skin behavior.

In the Glauber model the valence-nucleon's wave function can be
obtained with the three input parameters which has been mentioned
above. It is easy to get the density distribution from the wave
function after normalization. The solid line shown in Fig.4(c) is
the valence-proton of $^{29}$P density distribution from the
Glauber model. With the density distribution of the core,
$^{28}$Si, we get $^{29}$P density distribution, which is the
total one shown in Fig.4(c). Compared with the result from SHF
model, the density distribution of $^{29}$P from Glauber model is
suitable to embody the nuclear character and it would be reliable
to get the momentum distribution.

In summary, the experimental measurement of the parallel momentum
distribution of $^{28}$Si fragment from the breakup of 30.7
MeV/nucleon $^{29}$P has been reported for the first time. The
parallel momentum width is 110.5$\pm$ 23.5 MeV/c FWHM which is
consistent with the calculation of the Glauber model assuming the
core plus one proton. The Skyrme-Hartree-Fock calculation also
gives a proton-skin density distribution in comparison with
$^{28}$Si.  Taken together with the theoretical analysis of the
density distribution and the nucleon orbit arrangement, it is
obviously hard to form the halo structure in $^{29}$P. But the
possibility of the proton-skin structure of $^{29}$P  exists, and
this nucleus could be understood as the transition one between the
exotic and stable isotopes.

We appreciate helpful discussions with Prof. Gen-Ming Jin of
Institute of Modern Physics.

{}

\end{document}